\documentclass[pra,aps,a4paper,grouppedaddress,notitlepage,twocolumn,longbibliography]{revtex4-2}
\usepackage{amssymb}
\usepackage{amsmath}
\usepackage{amsfonts}
\usepackage{graphicx}
\usepackage{bm}
\usepackage[dvipsnames]{xcolor}
\usepackage{natbib}
\usepackage[bookmarksopen,colorlinks,linkcolor=blue,citecolor=Bittersweet,urlcolor=red]{hyperref}
\usepackage{esint}
\usepackage{dsfont}
\usepackage[title]{appendix}
\usepackage{comment}

\DeclareMathOperator{\tr}{tr}
\DeclareMathOperator{\erf}{erf}
\DeclareMathOperator{\erfc}{erfc}

\newcommand{\corr}[1]{\langle #1\rangle}
\newcommand{\ccorr}[1]{\langle\langle #1\rangle\rangle}

\newcommand{\Feff}{\tilde{\mathcal{F}}}

\def\be{\begin{equation}}
\def\ee{\end{equation}}

\def\bf{\mathbf{f}}

\def\br{\mathbf{r}}
\def\bR{\mathbf{R}}

\def\vp{\varphi}

\def\A{{\zeta_\alpha}}

\def\prof{g}

\begin{document}

\title{Statistics of the vortex pinning potential in superconducting films}

\author{Matvei S. Kniazev}
\affiliation{L.\ D.\ Landau Institute for Theoretical Physics, Chernogolovka 142432, Russia}
\affiliation{Moscow Institute of Physics and Technology, Dolgoprudny 141701, Russia}

\author{Nikolai A. Stepanov}
\affiliation{L.\ D.\ Landau Institute for Theoretical Physics, Chernogolovka 142432, Russia}
\affiliation{Moscow Institute of Physics and Technology, Dolgoprudny 141701, Russia}

\author{Mikhail A.\ Skvortsov}
\affiliation{L.\ D.\ Landau Institute for Theoretical Physics, Chernogolovka 142432, Russia}
\affiliation{Moscow Institute of Physics and Technology, Dolgoprudny 141701, Russia}

\date{\today}

\begin{abstract}
We investigate the statistical properties of the vortex pinning potential in a thin superconducting film. 
Modeling intrinsic inhomogeneities by a random-temperature Ginzburg-Landau functional with short-range Gaussian disorder, we derive the pinning landscape $E(\bR)$ by determining how the vortex core adapts to randomness. Within the hard-core approximation, applicable for weak disorder, the energy landscape exhibits Gaussian statistics.
In this regime, the mean areal density of its minima is given by $n_\text{min}\approx(6\xi)^{-2}$, indicating that the typical spacing between neighboring minima is significantly larger than the vortex core size $\xi$. Going beyond the hard-core approximation, we allow the vortex order parameter to relax in response to the inhomogeneities. As a result, the pinning potential statistics become non-Gaussian. We calculate the leading correction due to the core deformation, which reduces the density of minima with a relative magnitude scaling as $(T_c-T)^{-1/2}$.
\end{abstract}

\maketitle

\section{Introduction}

Vortex matter in superconductors exhibits remarkably complex behavior arising from the interplay of disorder, intervortex interaction, and fluctuations (both thermal and quantum), making it a field of sustained research interest \cite{Blatter1994}.
The motion of vortices driven by the transport current's Lorentz force leads to dissipative energy loss \cite{BS}, thereby undermining the defining property of a superconductor, its zero resistance. 
Therefore, high transport currents necessary for many practical superconducting applications can only be realized through the pinning of vortices by defects \cite{Si,Devoret,Welp,Gurevich}.

In 1970, Larkin \cite{Larkin-0} showed that inhomogeneities destroy the long-range order of the Abrikosov vortex lattice in the mixed state. 
This idea later evolved into the concept of collective pinning \cite{CollectivePinning}, where the elasticity of the vortex lattice transmits individual pinning forces throughout the entire array, leading to its immobilization as a whole.
A separate research direction focuses on how an impurity potential modifies quasiparticle states localized in the vortex cores in thin superconducting films. This ranges from mesoscopic effects on energy-level statistics for weak, point-like impurities \cite{FS97, LO98, Koulakov, SKF98, BHL99, Fujita} to notable changes in the mean density of states induced by extended defects \cite{Melnikov09, Melnikov2020, Khodaeva2022}.

Recent interest in vortex physics has been revived by advances in high-resolution experimental techniques that now allow the direct observation and tracking of individual vortices. A key breakthrough was the integration of scanning tunneling spectroscopy (STS) \cite{Suderow,Pratap2023} with nanoscale ultrasensitive SQUID interferometers, leading to imaging technologies such as SQUID-on-tip \cite{Zeldov-1,Zeldov-2,Zeldov2019,Poggio} and SQUID-on-lever \cite{Weber2025}. These advances have enabled researchers to measure the pinning force on individual vortices \cite{Zeldov-1}, follow superfast vortex motion under high current densities \cite{Zeldov-2}, observe thermally activated vortex hopping \cite{Poggio}, and obtain magnetic images of single skyrmions \cite{Weber2025}.

Complementary techniques have further expanded experimental capabilities. The magnetic field distribution of vortices has been measured with high sensitivity using diamond nitrogen-vacancy center magnetometry \cite{Thiel2016}. Time-resolved nanothermometry has been used to quantify energy dissipation during vortex motion \cite{Foltyn2024}, while time-resolved STS imaging has revealed a highly inhomogeneous state of vortex lattice with coexisting pinned and mobile vortices \cite{Pratap2023}.

By using a thin superconducting film with a small penetration depth $\lambda\sim\xi$, Embon \emph{et al}.\ \cite{Zeldov-1} suppressed vortex-vortex interactions, enabling the study of an individual vortex, which behaves as a classical particle in a random potential landscape $E(\bR)$. By tracking its trajectory under a varying Lorentz force from an applied current, they directly characterized the nanoscale two-dimensional (2D) pinning potential.
Sample regions where a vortex is stably trapped at various driving currents correspond to positive curvatures of $E(\bR)$. This links pancake vortex pinning to the study of the Hessian stability map \cite{Bray2007,Yan2018} of a random function on a plane.
In the vortex context, this problem was recently studied for several types of the pinning potential \cite{Geshkenbein2022,Geshkenbein2023}. In the case of a Gaussian random field, the stable area fraction was found to be $(3-\sqrt{3})/6\approx 21\%$.

In Ref.\ \cite{Geshkenbein2022,Geshkenbein2023}, a vortex was treated as a simple 2D particle in a given random potential. In reality, a vortex (even pancake) is an extended object, possessing multiple internal degrees of freedom, and the effective potential it experiences must be derived from a specific microscopic model of inhomogeneity.
In the theory of pinning, disorder is commonly incorporated phenomenologically by introducing frozen-in spatial fluctuations of the parameters in the Ginzburg-Landau (GL) functional \cite{CollectivePinning,Larkin-0,Larkin-1}. Those are typically dealt with in the hard-core approximation, where the vortex’s order‑parameter profile is taken to be unaffected by disorder. In this approach, the randomness in the GL coefficients is effectively smoothed over the core size $\xi$, which produces an effective pinning landscape $E(\bR)$ for the vortex‑center coordinate $\bR$.

In this paper, we develop a statistical description of the vortex pinning potential in thin superconducting films in the vicinity of the critical temperature $T_c$. Starting with the random-temperature GL model with Gaussian short-range correlations, we derive the pinning energy landscape $E(\bR)$ and compute the areal density of its extremal points. In the hard-core approximation applicable for weak disorder, the density of minima is 
\be
\label{n-min-res}
  n_\text{min}
  =
  \frac{0.0275}{\xi^2},
\ee
where $\xi$ is the temperature-dependent coherence length.
The surprisingly small numerical factor in Eq.\ \eqref{n-min-res} implies that the characteristic spacing between pinning minima, $n_\text{min}^{-1/2}\approx 6\xi$, is much larger than the vortex core size $\xi$.

Going beyond the hard‑core approximation, we include the adjustment of the vortex order‑parameter profile in response to inhomogeneities. Rather counterintuitively, this effect reduces $n_\text{min}$, with the relative correction scaling as $\sqrt{\tau_*/\tau}$, where $\tau=1-T/T_c$. 
Here $\tau_*$ is the dimensionless inhomogeneity strength,
that sets the width of the temperature region near $T_c$ where quenched fluctuations alter the nature of the transition \cite{Zuev}.

This paper is organized as follows. In Sec. \ref{S:Model}, we introduce the model, connect the density of extrema to the joint distribution of the local force and Hessian matrix, and discuss the inherent ambiguity in defining the vortex position.
The density of extrema and curvature distribution in the hard-core limit are calculated in Sec.\ \ref{S:HardCore}. This approximation is relaxed in Sec.\ \ref{S:SoftCore}, where we perturbatively account for vortex-profile deformation due to disorder and compute the resulting correction to the density of extrema. 
Section \ref{S:Conclusion} provides a short summary.
Methodological and computational aspects are detailed in Appendices.

\section{Model and basic relations}
\label{S:Model}

\subsection{Random-temperature Ginzburg-Landau model}

An inhomogeneous superconductor  in the vicinity of the critical temperature $T_c$ is described by the Ginzburg-Landau free energy functional for the complex order parameter $\Delta(\br)$ given by \cite{Gorkov}
\begin{equation}
\label{def-GL-Functional}
    {\cal{F}}
    =
    \int d\br
    \left\{
      \gamma |\nabla\Delta|^2 
    - [\alpha+\delta\alpha(\br)]|\Delta|^2
    + \frac{\beta}2 |\Delta|^4 
    \right\} .
\end{equation}
The coefficient $\alpha=\nu(1-T/T_c)$ is positive in the superconducting state, changing sign at the transition. 
The equilibrium value of the order parameter is given by $\Delta_0 = \sqrt{\alpha/\beta}$. Spatial rigidity of the superconductor is characterized by the coherence length $\xi=\sqrt{\gamma/\alpha}$.
In the mean-field approximation, the order parameter $\Delta(\br)$ satisfies the GL equation
\begin{equation}
\label{GLE}
    -\gamma \nabla^{2}\Delta
    -[\alpha+\delta\alpha(\br)]\Delta
    +\beta\left|\Delta\right|^2\Delta
    =
    0 .
\end{equation}

The spatial inhomogeneity of the material is modeled phenomenologically through quenched Gaussian fluctuations in $\alpha$ \cite{CollectivePinning,Larkin-1,Zuev}, which are taken to be short-ranged with the correlation function
\be
\label{da-da}
  \corr{\delta\alpha(\br)\delta\alpha(\br')}
  %=\A\,\delta(\br-\br')
  = \A \, \delta(\br-\br') .
\ee

Assuming the film is sufficiently thin, we treat the system as two-dimensional with $\br=(x,y)$. The strength of the inhomogeneities is then characterized by a dimensionless parameter
\be
\label{tau*}
  \tau_* =
  \frac{\A}{4\pi\gamma\nu} ,
\ee
introduced in Refs.\ \cite{SF2005,Zuev} as an analog of the Ginzburg number (in Ref.\ \cite{SF2005}, it was referred to as $\delta_d$). It defines the temperature window of size $\delta T\sim T_c\tau_*$ near $T_c$, where frozen irregularities become strong and change the nature of the transition.

Equation \eqref{def-GL-Functional} completely neglects magnetic field effects by assuming the extreme type-II limit, with the penetration length $\lambda\gg\xi$. For thin films with thickness $d\ll\lambda$, this condition becomes even less restrictive: $\Lambda\gg\xi$, where
$\Lambda=\lambda^2/d$ is the Pearl length \cite{Pearl}.

\begin{figure}[b]
\includegraphics[width=0.95\linewidth]{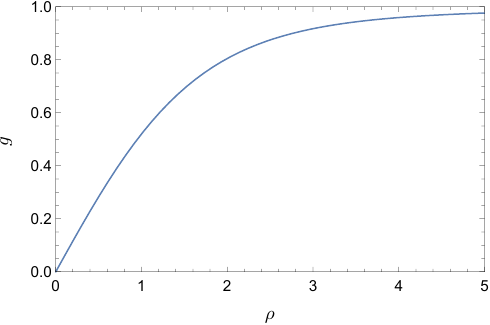}
\caption{Radial profile of the vortex order parameter $\prof(\rho)$ in a homogeneous system, see Eq.\ \eqref{vortex-g}.}
\label{F:radial}
\end{figure}

In a homogeneous system, the vortex solution located at the coordinate center can be written as
\be
\label{vortex-g}
  \Delta_0(\br) = \Delta_0 \prof(r/\xi) e^{i\vp},
\ee
where $\vp$ is the polar angle. The radial function $\prof(\rho)$ satisfies the nonlinear differential equation
\begin{equation}
    \prof''
    +\frac{\prof'}{\rho}
    -\frac{\prof}{\rho^2}
    +\prof-\prof^{3}=0 ,
\end{equation}
with the boundary conditions $\prof(0)=0$ and $\prof(\infty)=1$. Its solution \cite{Abrikosov} is shown in Fig.\ \ref{F:radial}.
The vortex energy logarithmically diverges at large scales, where it should be cut at the penetration length: $E_0\approx2\pi\gamma\Delta_0^2\ln(\Lambda/\xi)$.
As shown below, vortex energy corrections in the presence of inhomogeneities originate from the region $r\sim\xi$, thus justifying the neglect of magnetic terms in Eq.\ \eqref{def-GL-Functional} for the vortex pinning problem.

\subsection{Density of stable points}
\label{S:Density of stable points}

In a homogeneous system, a single vortex with the order parameter $\Delta_0(\br-\bR)$ can be located everywhere, and its energy is independent of $\bR$ (zero mode). Inhomogeneity in the coefficient $\delta\alpha(\br)$ breaks translational invariance and makes the vortex energy $E(\bR)$ an explicit function of its position $\bR$. In the absence of an external force, either from the transport current or from other vortices, a vortex will find itself in a minimum of the potential $E(\bR)$. An important characteristic of the random landscape $E(\bR)$ is the number of its minima per unit area, $n_\text{min}$. This is a crude quantity that merely counts all vortex pinning centers, both deep and shallow. A more detailed information is provided by the curvature distribution at the minima.

Once the function $E(\bR)$ is known, the total number of its extrema can be calculated as
\be
\label{Ntot}
  N_\text{tot}
  =
  \int dx\,dy\,
  \delta(\bf)
  \,|\!\det H| ,
\ee
where $\bf$ is the (minus) force acting on the vortex:
\be
  f_{i}(\bR)
  =
  \frac{\partial E(\bR)}{\partial R_i} ,
\ee
and $H$ is the Hessian matrix with the elements
\be
\label{hessian-def}
  h_{ij}(\bR)
  =
  \frac{\partial^2 E(\bR)}{\partial R_i \partial R_j}
%  \begin{pmatrix}
%    f_{xx} & f_{xy}\\
%    f_{yx} & f_{yy}
%  \end{pmatrix}
  .
\ee

The eigenvalues of $H$ are the two principal curvatures, $\kappa_1$ and $\kappa_2$, of the potential landscape $E(\bR)$ at extremal points. Their signs determine the type of the stable point: 
a minimum ($\kappa_{1,2}>0$), a maximum ($\kappa_{1,2}<0$), or a saddle ($\kappa_1\kappa_2<0$).

Averaging Eq.\ \eqref{Ntot} over inhomogeneities, we obtain an expression for the mean density of extrema:
\be
\label{n-tot-def}
  n_\text{tot}
  =
  \corr{\delta(\bf) | h_{xx}h_{yy}-h_{xy}^2 |} .
\ee
To determine the density of a particular type of extrema ($a=$ min, max, sad), one should add a corresponding counting condition:
\be
\label{n-a-def}
  n_a
  =
  \corr{\Theta_a \delta(\bf) | h_{xx}h_{yy}-h_{xy}^2 |}
  ,
\ee
where $\Theta_\text{min}=\theta(\kappa_1)\theta(\kappa_2)$,
$\Theta_\text{max}=\theta(-\kappa_1)\theta(-\kappa_2)$,
and $\Theta_\text{sad}=\theta(-\kappa_1\kappa_2)$.

We combine the five local derivatives of $E(\bR)$ appearing in Eq. \eqref{n-a-def} into a 5-vector
\be
\label{def-comp-f}
  \mathfrak{f} = 
  (f_x, f_y, h_{xx}, h_{xy}, h_{yy}) 
\ee
and write its components in a concise form
\be
\label{dmu}
  \mathfrak{f}_\mu(\bR)
  = 
  \hat\partial_\mu E(\bR) ,
\ee
where
\be
  \hat\partial_\mu
  =
  ( \partial_{R_x} , \partial_{R_y} , \partial_{R_x}^2 ,
    \partial_{R_x}\partial_{R_y} , \partial_{R_y}^2 )
  .
\ee
Hereafter Greek indices take the values $x,\,y,\,xx,\,xy,\,yy$.

Ensemble averaging in Eq.\ \eqref{n-a-def} can be easily performed, provided the joint distribution function $P(\mathfrak{f})$ is known. Then averaging over a random field $\delta\alpha(\br)$ is equivalent to averaging over a five-component random vector $\mathfrak{f}$ with the distribution function $P(\mathfrak{f})$. Hence the problem reduces to finding $P(\mathfrak{f})$ for a given inhomogeneity statistics \cite{Rice1,Rice2}.

\subsection{Perturbative determination of the vortex energy}
\label{S-Perturbation approach}

For weak inhomogeneity, the position-dependent vortex energy $E(\bR)$ can be determined perturbatively. To this end, we write the order parameter as
\be
\label{Delta-exp}
  \Delta(\br) 
  =
  \Delta_0(\br-\bR) + \Delta_1(\br) + \Delta_2(\br) + \dots ,
\ee
with $\Delta_n(\br)$ scaling as the $n$th power of $\delta\alpha$, and arrange the term $\Delta_n$ and its complex conjugate into a single vector $\hat{\Delta}_n=(\Delta_n,\Delta^*_n)$.
Substituting the series \eqref{Delta-exp} into the GL Eq.\ \eqref{GLE}, one can obtain $\hat\Delta_n$ iteratively following the approach developed in Ref.\ \cite{Zuev}. In particular, the first-order correction satisfies the linear equation
\be
\label{Delta-1-eq}
\hat{L}^{-1}
\hat{\Delta}_1
=
  \delta\alpha(\br)
  \hat{\Delta}_0
  ,
\ee
with the differential operator in the left-hand side,
\be
\label{iL-def}
\hat{L}^{-1}
=
\begin{pmatrix}
    -\gamma \nabla^2-\alpha+2\beta|\Delta_0|^2 & \beta\Delta_0^2 \\
    \beta\left(\Delta_0^*\right)^2 &  -\gamma \nabla^2-\alpha+2\beta|\Delta_0|^2 
  \end{pmatrix}
  ,
\ee
known as the inverse fluctuation propagator.

Plugging Eq.\ \eqref{Delta-exp} into the GL free energy \eqref{def-GL-Functional}, we obtain an associated perturbative expansion for the vortex energy in powers of $\delta\alpha$:
\be
\label{Energy series}
  E(\bR) = E_0 + E_1(\bR) + E_2(\bR) + \dots 
\ee
The position-dependent parts of the vortex energy 
calculated in Appendix \ref{Appendix-A} have the form
\begin{subequations}
\label{E12}
\begin{gather}
\label{E1}
  E_1(\bR)
  =
  - \int d\br\,\delta\alpha(\br) 
    \left|\Delta_0\right|^2
  ,
\\
\label{E2}
  E_2(\bR)
  =
   - \frac{1}{2}
  \int d\br\,\delta\alpha(\br)\,
  \hat{\Delta}^\dagger_0\hat{\Delta}_1
    .
\end{gather}
\end{subequations}

Taking the derivatives of Eq.\ \eqref{Energy series} with respect to the vortex coordinate $\bR$, we get a similar series for the 5-vector $\mathfrak{f}$:
\be
\label{Force series}
  \mathfrak{f}_\mu 
  = 
  \mathfrak{f}^{(1)}_\mu 
+ \mathfrak{f}^{(2)}_\mu + \dots 
\ee
Using Eqs.\ \eqref{E12}, we immediately obtain 
\begin{subequations}
\label{f12}
\begin{gather}
\label{f-exp-1ord}
  \mathfrak{f}_\mu^{(1)}
  =
  \int d\br\,\mathfrak{b}_\mu(\br;\bR)\delta\alpha(\br)
  ,
\\
  \label{f-exp-2ord}
\mathfrak{f}_\mu^{(2)}
  =
  \frac{1}{2}\int d\br \, d\br'
  \mathfrak{c}_\mu(\br,\br';\bR)\delta\alpha(\br)\delta\alpha(\br') ,
\end{gather}
\end{subequations}
with the kernels
\begin{subequations}
\label{Def-b-c}
\begin{gather}
\label{def-b}
  \mathfrak{b}_\mu(\br;\bR)
  =
  - \hat\partial_\mu 
  |\Delta_{0,\bR}(\br)|^2 
  ,
\\
  \label{def-c}
 \mathfrak{c}_\mu(\br,\br^\prime;\bR)
 =
 -\hat\partial_\mu \big[
 \hat{\Delta}_{0,\bR}^\dagger(\br)\hat{L}_\bR(\br,\br')\hat{\Delta}_{0,\bR}(\br') \big].
  %  -\hat\partial_\mu \big[
  % \Delta_0^*(\br-\bR) \hat{L}(\br,\br';\bR)\Delta_0(\br'-\bR)
  % \big]
  % +\mathrm{c.c.} ,
\end{gather}
\end{subequations}
where $\Delta_{0,\bR}(r) \equiv \Delta_0(\br-\bR)$.

Note that the lowest-order correction $E_1(\bR)$ is obtained just by filtering $\delta\alpha(\br)$ with the \emph{unperturbed} vortex profile. This simplest case of a \emph{hard-core} vortex will be considered in Sec.\ \ref{S:HardCore}. It is applicable for sufficiently weak inhomogeneities, whereas going beyond the first order requires recalculating the shape of the vortex adapting to the disorder potential, see Sec.\ \ref{S:SoftCore}.

\subsection{Not invertible  $\hat L^{-1}$,  zero modes, and the vortex center definition}
\label{SS:Disclaimer}

While intuitive, the perturbative approach outlined in Sec.\ \ref{S-Perturbation approach} suffers from fundamental conceptual issues. 
This problem arises already at the level of the first correction $\Delta_1$, since the operator $\hat{L}^{-1}$ in Eq.\ \eqref{Delta-1-eq} is not invertible. Mathematically, the problem stems from the existence of its zero modes $|0_x\rangle$ and $|0_y\rangle$ with the eigenfunctions
\be
\label{def-zero-mode}
  \hat\psi_{x,y}
  =
  \corr{\br|0_{x,y}}
  =
  \partial_{R_{x,y}} \hat{\Delta}_0(\br-\bR) ,
\ee
which are associated with the vortex translations in a homogeneous system.
(In the absence of a vortex, the spectrum of $\hat L^{-1}$ is gapped and the perturbation theory in $\delta\alpha$ is well-defined \cite{Zuev}.)

An attempt to project the right-hand side of Eq. \eqref{Delta-1-eq} onto the subspace orthogonal to $|0_{x,y}\rangle$ also encounters technical obstacles related to normalization of the zero modes \eqref{def-zero-mode}. Indeed, for the 2D vortex problem considered, the square of their norm $\corr{0_i|0_i} = \int\hat\psi_i^\dagger\hat\psi_i \, d^2r$ ($i=x$ or $y$) logarithmically diverges at large distances. 
Cutting the logarithm off at the Pearl length $\Lambda$ seems appropriate, however, this modification necessarily implies that the vector potential must be retained in the initial GL equation \eqref{GLE}.

The mathematical ill-posedness of the problem is a counterpart of the physical ambiguity in the notion of vortex center in the presence of defects. While the vortex center $\bR$ coincides with the zero of $\Delta(\br)$ in a clean system, its identification in an inhomogeneous environment is no longer unique. A particularly simple and convenient operational definition of the functional $\bR[\Delta(\br)]$ comes from the requirement that it minimizes the weighted quadratic deviation from the bare vortex solution,
\be
\label{W-exp}
  W[\Delta(\br);\bR]
%  W_\Upsilon[\Delta(\br);\bR]
  =
  \int d\br\, \Upsilon(|\br-\bR|) |\Delta(\br)-\Delta_{0,\bR}(\br)|^2 ,
\ee
specified by a particular profile $\Upsilon(r)$. The resulting minimization problem with the constraint $\partial W/\partial\bR=0$, should be solved by the method of Lagrange multipliers.

With the most straightforward choice $\Upsilon=1$ adopted in Refs.\ \cite{Zittartz,Larkin-2}, the right-hand side of Eq.\ \eqref{Delta-1-eq} acquires an additional term $\lambda_x\hat\psi_x+\lambda_y\hat\psi_y$. The proper choice of Lagrange multipliers $\lambda_x$ and $\lambda_y$ should ensure orthogonality of $\delta\alpha(\br)\hat{\Delta}_0+\lambda_x\hat\psi_x+\lambda_y\hat\psi_y$ and the zero modes. However, due to the above-mentioned logarithmic divergency of the norm $\corr{0_i|0_i}$, such an approach cannot be rigorously justified for the pancake vortex.

The introduction of a decaying function, e.g., $\Upsilon(r) = \exp(-r^2/\Lambda^2)$, regularizes the divergent normalization of the zero-mode, making the mathematical problem well-posed. 
However, the term $\partial\Upsilon(|\br-\bR|)/\partial\bR$ would then modify the operator $\hat L^{-1}$ in Eq.\ \eqref{Delta-1-eq}.
On the other hand, since we have already neglected magnetic field effects in the free energy \eqref{def-GL-Functional}, we can safely omit this extra term without loss of accuracy, provided $\Lambda$ is of the order of the Pearl length.

Applying the Lagrange multiplier method replaces $\hat{L}$
in Eq.\ \eqref{Delta-1-eq} with a modified operator 
$\hat L_\text{reg}$, effetively projecting onto the subspace orthogonal to the zero modes, see Appendices~\ref{Appendix-zero-mode} and \ref{Appendix-J-I}.

Thus the definition of the vortex center $\bR[\Delta(\br)]$ is noninvariant and inevitably depends on the particular choice of the regularizing function $\Upsilon(r)$. Also noninvariant (beyond the first order) is the vortex energy landscape $E(\bR)$ and related characteristics: the force $f_i(\bR)$, the Hessian matrix $h_{ij}(\bR)$, and their distribution function $P(\mathfrak{f})$. At the same time, the total number of various types of $E[\bR]$ extrema should be $\Upsilon$-independent, as they are related to the exact solution of the GL equation \eqref{GLE} for $\Delta(\br)$.

Since we are interested only in such invariant quantities, we will formally proceed with the ill-defined operator $\hat L$ instead of the regularized $\hat L_\text{reg}$. As we will see in Sec.\ \ref{Sec-density-ext}, the structure of the resulting expressions for the average density of extrema automatically projects onto the zero-mode-free sector, which \emph{a posteriori} justifies the brute-force approach.

\section{Hard core}
\label{S:HardCore}

\subsection{Distribution function $P_0(\mathfrak{f})$}

For superconducting films exhibiting sufficiently weak inhomogeneity, one can use the first-order expression \eqref{E1} for the vortex energy, as it was done in Ref. \cite{Larkin-1}. In this approximation, the vector $\mathfrak{f}=\mathfrak{f}^{(1)}[\delta\alpha]$ is a linear functional of $\delta\alpha$ given by Eq.\ \eqref{f-exp-1ord}.
Since the distribution of $\delta\alpha(\br)$ was assumed to be Gaussian with zero mean, the distribution $P_0(\mathfrak{f})$ is also Gaussian with zero mean and has a generic form
\be
\label{f-dist-1ord}
  P_0(\mathfrak{f})
  =
  \frac{\sqrt{\det B}}{\left(2\pi\right)^{5/2}}
  \exp\left(
    - \frac{\mathfrak{f}_\mu B_{\mu\nu}\mathfrak{f}_\nu}{2}
  \right)
\ee
with some $5\times5$ matrix $B$. The latter is completely specified by the correlation functions $\corr{\mathfrak{f}_\mu\mathfrak{f}_\nu}$.
For the white-noise disorder in $\alpha$ [see Eq.~\eqref{da-da}], the forces $f_{i}$ ($i=x,\,y$) and the matrix elements of the Hessian $h_a$ 
($a=xx,\,xy,\,yy$)
are independent, with nonzero correlators given by
\begin{subequations}
\label{corrs}
\begin{gather}
  \corr{f_i^2}
  =
  c_1 \A\Delta_0^4 ,
\\
\label{corrs2}
  \corr{h_ah_b}
  =
  c_2\frac{\A \Delta^4_0}{\xi^2}
  C_{ab} .
\end{gather}
\end{subequations}
Here the matrix $C$ is defined as
\be
  C 
  =
  \begin{pmatrix}
    3 & 0 & 1\\
    0 & 1 & 0\\
    1 & 0 & 3
  \end{pmatrix} ,
\ee
where the coefficient 3 arises from the ratio of angular averages $\corr{n_x^4}/\corr{n_x^2n_y^2}$ for a two-dimensional unit vector $(n_x,n_y)$.
The form of the matrix $C$ is completely determined by the assumption of Gaussian statistics \cite{Geshkenbein2022}.
The coefficients $c_1$ and $c_2$ are expressed in terms of the dimensionless vortex profile function $\prof(\rho)$, introduced in Eq.\ \eqref{vortex-g}, as
\begin{subequations}
\label{c1c2}
\begin{gather}
\label{c1}
  c_1
  =
  % 4\pi\int d\rho\, \rho \, \prof^2 \prof'^2
  % \approx 1.24 ,
  % \\
   \pi\int d\rho\, \rho \left(\frac{\partial \prof^2}{\partial\rho}\right)^2
  \approx 1.24 ,
\\
\label{c2}
    c_2
    =
    \frac{\pi}{4} \int d\rho\, \rho \left(
    \frac{1}{\rho}\frac{\partial}{\partial\rho}\rho\frac{\partial}{\partial\rho}\prof^2
    \right)^2
    \approx
    0.371 .
    % \\
    % \pi \int d\rho\, \rho (\prof\prof''+\prof'^2-\prof\prof'/\rho)^2
    % \approx
    % 0.371 .
\end{gather}
\end{subequations}
Note that while the total vortex energy is logarithmic, coming from the interval $\xi<r<\Lambda$, the integrals \eqref{c1c2} converge at the vortex core size, $r\sim\xi$.

Equations \eqref{corrs} and \eqref{c1c2} agree with results of Larkin and Ovchinnikov for $\corr{(\nabla E)^2}$ and $\corr{(\nabla^2 E)^2}$ \cite{Larkin-1} (with our $c_i$ related to their $I_j$ via $c_1=\pi I_1$ and $c_2=\pi I_3/4$).

Inverting the matrix $\corr{\mathfrak{f}_\mu\mathfrak{f}_\nu}$, 
we obtain $B$ in the form
\be
  B 
  = 
  \begin{pmatrix}
    B^{(1)} & 0\\
    0 & B^{(2)}
  \end{pmatrix},
\ee
with the $2\times2$ block $B^{(1)}$ describing the distribution of the force $(f_x,f_y)$, and the $3\times3$ block $B^{(2)}$ describing the distribution of Hessian $(h_{xx},h_{xy},h_{yy})$:
\be
  B^{(1)}
  =
  \frac{1}{c_1\A\Delta_0^4}
  \begin{pmatrix}
    1 & 0\\
    0 & 1
  \end{pmatrix},
\qquad
  B^{(2)}
  =
  \frac{\xi^2}{c_2\A\Delta_0^4} C^{-1} .
\ee

\subsection{Density of extrema}
\label{Sec:Density of extrema}

Once the distribution function $P_0(\mathfrak{f})$ is known, one can readily determine the density of stable points. We start with calculating the total density given by Eq.~\eqref{n-tot-def}. Due to the presence of $\delta(\mathbf{f})=\delta(f_x)\delta(f_y)$ it remains to integrate over the three independent elements of the Hessian matrix, $h_a$ ($a=xx,\,xy,\,yy$):
\be
\label{ntot1}
  n_\text{tot}
  =
  \frac{\sqrt{\det B}}{(2\pi)^{5/2}}
  \int d^3h_a 
  \left|h_{xx}h_{yy}-h_{xy}^2\right| 
  \exp\biggl(-\frac{h_a B^{(2)}_{ab}h_b}{2}\biggr) .
\ee
Switching to dimensionless variables $t_a$ introduced as $t_a=\sqrt{\xi^2/2c_2\A\Delta_0^4}\,h_a$ and computing $\det B$, we arrive~at
\be
  n_\text{tot}
  =
  \frac{c_\text{tot}c_2}{\pi c_1\xi^2}
  ,
\ee
where the coefficient $c_\text{tot}$ is determined by the triple integral
\be
\label{I-tot-App}
  c_\text{tot}
  =
  \int \frac{d^3t_a}{(2\pi)^{3/2}}
  \left|t_{xx}t_{yy}-t_{xy}^2\right|
  \exp\bigl(-t_a C^{-1}_{ab}t_b\bigr) .
\ee
To calculate it, we write $|t_{xx}t_{yy}-t_{xy}^2|$ via the Fourier transform, using
\be
\label{mod-FT}
  |z|
  =
  \int\frac{d\lambda}{2\pi} m(\lambda) \exp\left(i\lambda z\right)
\ee
with $m(\lambda)=-1/(\lambda-i0)^2-1/(\lambda+i0)^2$. This renders the $t_a$ integration Gaussian, leading to
\be
\label{Itot2}
  c_\text{tot}
  =
  \int\frac{d\lambda}{2\pi}
  \frac{m(\lambda)}{\sqrt{\det(1-i\lambda CD)}} ,
\ee
where we introduced an auxiliary matrix
\be
  D 
  = 
  \begin{pmatrix}
    0 & 0 & 1/2\\
    0 & -1 & 0\\
    1/2 & 0 & 0
  \end{pmatrix} .
\ee
Evaluating the determinant, we obtain
\be
  c_\text{tot}
  =
  \int\frac{d\lambda}{2\pi}
  \frac{m(\lambda)}{(1+i \lambda) \sqrt{1-2i\lambda}}
  =
  \frac{2}{\sqrt3} ,
\ee
where the result is determined by the residue at the pole $\lambda=i$. [Note that the pole of $m(\lambda)$ at $\lambda=+i0$ does not contribute, as the corresponding residue vanishes due to $\tr CD=0$, which is equivalent to the identity $\corr{h_{xx}h_{yy}} = \corr{h_{xy}^2}$, see Eq.\ \eqref{corrs2}.] 
Hence we get for the total density of extrema:
\begin{equation}
 \label{n-tot-linear}
  n_\text{tot}
  =
  \frac{2c_2}{\sqrt{3}\pi c_1}\frac{1}{\xi^2}
  .
\end{equation}

For the Gaussian statistics of $\mathfrak{f}$ applicable in the hard-core limit, the density of minima, maxima, and saddles is 1/4, 1/4, and 1/2 of $n_\text{tot}$, respectively. Indeed, $n_\text{min}+n_\text{max}$ can be written in the form of Eq.\ \eqref{ntot1} with an additional factor of $\Theta_\text{min}+\Theta_\text{max} = \theta(h_{xx}h_{yy}-h_{xy}^2)$. Repeating the derivation, we arrive at an analog of Eq.\ \eqref{Itot2}, with $m(\lambda)$ replaced by $m(\lambda)=-1/(\lambda-i0)^2$. Since the pole at $+i0$ is effectively annihilated by the term $\det(1-i\lambda CD)$, as discussed above, the result for $n_\text{min}+n_\text{max}$ is just $n_\text{tot}/2$. Finally, the symmetry $P_0(\mathfrak{f})=P_0(-\mathfrak{f})$ ensures that $n_\text{min}=n_\text{max}$. Hence, $n_\text{min}=n_\text{tot}/4$, and we arrive at Eq.\ \eqref{n-min-res}.

\subsection{Local and global curvature distributions}

Our calculation of $P_0(\mathfrak{f})$ enables us to investigate the curvature distribution function. 
%$P(\kappa_1,\kappa_2)$, where $\kappa_{1,2}$ are the Hessian eigenvalues. 
This distribution can be defined either \emph{globally} (at any point in the landscape) or \emph{locally} (restricted to extremal points). The global case has recently been studied in the context of Hessian stability maps \cite{Geshkenbein2022}.

We start with defining the global and local distributions of the principal curvatures $\kappa_{1,2}$ (the Hessian eigenvalues):
\begin{subequations}
\begin{gather}
\label{P-k1-k2-global-def}
  P_\text{global}(k_1,k_2)
  =
  \corr{\delta(k_1-\kappa_1)\delta(k_2-\kappa_2)}
  ,
\\
\label{P-k1-k2-def}
  P_a(k_1,k_2)
  =
  \corr{\delta(k_1-\kappa_1)\delta(k_2-\kappa_2)
  \Theta_a \delta(\bf) 
  |{\det H}|}/{n_a}
  ,
\end{gather}
\end{subequations}
where, as usual, the index $a\in\{\text{min, max, sad}\}$.
A simple averaging with respect to $P_0(\mathfrak{f})$ given by Eq.\ \eqref{f-dist-1ord} yields
\begin{gather}
  P_\text{global}(\kappa_1,\kappa_2)
  =
  \frac{|\kappa_1-\kappa_2|}{2\sqrt{\pi}\,k_0^3}
  e^{-\kappa_i\Omega_{ij}\kappa_j/2} ,
\\
\label{PDF-cur-1-2}
    P_a(\kappa_1,\kappa_2)=
    \Theta_a
    \sqrt{\frac{3}{\pi}}
    \frac{n_\text{tot}}{2n_a}
    \frac{|\kappa_1\kappa_2(\kappa_1-\kappa_2)|}{k_0^5}
    e^{-\kappa_i\Omega_{ij}\kappa_j/2} ,
\end{gather}
where 
 $k_0^2=4c_2\A\Delta_0^4/\xi^2$ and
\begin{equation}
   \Omega
   =
   \frac{1}{2k_0^2}
   \begin{pmatrix}
      3 & -1\\
      -1 & 3
   \end{pmatrix}
   .
\end{equation}

\begin{figure}
\includegraphics[width=0.95\linewidth]{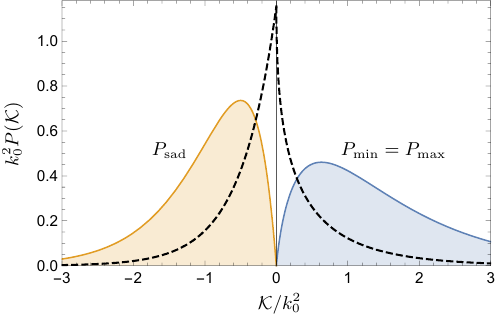}
\caption{Global (dashed) and local (solid, filled) Gaussian curvature probability distribution functions for the vortex energy landscape $E(\bR)$ in the hard-core approximation, as given by Eqs.\ \eqref{curvK}.}
\label{F:curvK}
\end{figure}

The distribution functions of the Gaussian curvature $\mathcal{K}=\det H=\kappa_1\kappa_2$ take the form
\begin{subequations}
\label{curvK}
\begin{gather}
\label{PglobK}
  P_\text{global}({\cal K})
  =
  \frac{2}{\sqrt{3}k_0^2}
  e^{2 {\cal K}/k_0^2}
  \left[ 
    1 - \theta({\cal K}) \erf\sqrt{\frac{3{\cal K}}{k_0^2}} \,
  \right] ,
\\
\label{Gauss curvature PDF}
  P_\text{min}(\mathcal{K})
  =
  P_\text{max}(\mathcal{K})
  =
  \theta(\mathcal{K})
  \frac{4\mathcal{K}}{k_0^4}e^{2\mathcal{K}/k_0^2}
  \erfc\sqrt{\frac{3\mathcal{K}}{k_0^2}} ,
\\
  P_\text{sad}({\cal{K}})
  =
  \theta(-{\cal{K}})\frac{4|{\cal{K}}|}{k_0^4}e^{2{\cal{K}}/k_0^2} .
\end{gather}
\end{subequations}
Equation \eqref{PglobK} has been derived in Ref.\ \cite{Geshkenbein2022}. The equivalence is established through the relation $G^{(4)}_0=\corr{h_{xx}h_{yy}}=c_2\A\Delta_0^4/\xi^2$.

The distribution functions of the mean curvature $\mathcal{H}=\tr H/2=(\kappa_1+\kappa_2)/2$ are given by
\begin{subequations}
\label{curvH}
\begin{equation}
   P_\text{global}({\cal{H}})
   =
   \frac{1}{\sqrt\pi \, k_0} e^{-{\cal H}^2/k_0^2} ,
\end{equation}
\vspace{-10pt}
\begin{multline}
    P_\text{min}(\mathcal{H})
    =
    P_\text{max}(-\mathcal{H})
\\{}
    =
    \frac{\theta(\mathcal{H})}{k_0}
    \sqrt{\frac{12}{\pi}}
    \left[
    e^{-\mathcal{H}^2/k_0^2}\left(\frac{2\mathcal{H}^2}{k_0^2}-1\right)
    +
    e^{-3\mathcal{H}^2/k_0^2}
    \right] ,
\end{multline}
\vspace{-10pt}
\be
   P_\text{sad}({\cal{H}})
   =
   \sqrt{\frac{3}{\pi}}
   \frac{1}{k_0}
   e^{-3\mathcal{H}^2/k_0^2} .
\ee
\end{subequations}

\begin{figure}
\includegraphics[width=0.95\linewidth]{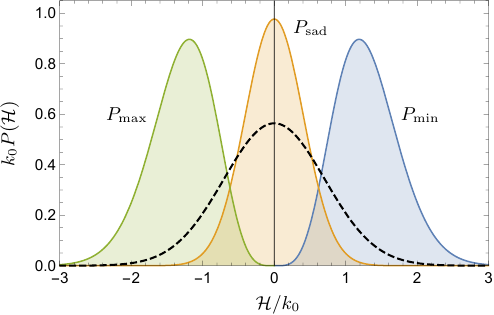}
\caption{Global (dashed) and local (solid, filled) mean curvature probability distribution functions for the vortex energy landscape $E(\bR)$ in the hard-core approximation, as given by Eqs.\ \eqref{curvH}.}
\label{F:curvH}
\end{figure}

The resulting distribution functions are presented in Figs.\ \ref{F:curvK} and \ref{F:curvH}. Due to an additional factor of $|{\det H}|$ in Eq.\ \eqref{P-k1-k2-def}, local distributions of Gaussian curvature are forced to vanish at zero $\mathcal{K}$. Accordingly, the curvature at minima is typically much larger than at an arbitrary point in the landscape.

\section{Soft core}
\label{S:SoftCore}

\subsection{Distribution function $P(\mathfrak{f})$}

Here we go beyond the hard core approximation developed in Sec.\ \ref{S:HardCore} and consider the effects of the order parameter modification by disorder. In the lowest order, they are contained in the term $\Delta_1$ in Eq.\ \eqref{Delta-exp}, which results in the quadratic-in-$\delta\alpha$ correction $\mathfrak{f}^{(2)}$ given by Eq.\ \eqref{f-exp-2ord}. The distribution function for the 5-vector $\mathfrak{f}$ then can be written as
\begin{equation}
\label{P20}
  P_{2}(\mathfrak{f})
  =
  \bigl<
  \delta\bigl(
    \mathfrak{f}
  - \mathfrak{f}^{(1)}[\delta\alpha]
  - \mathfrak{f}^{(2)}[\delta\alpha]
  \bigr)
  \bigr>,
\end{equation}
where Gaussian averaging over $\delta\alpha(\br)$ with the white-noise correlator \eqref{da-da} should be performed.
Writing Eq.\ \eqref{P20} via the generating function and expanding in $\mathfrak{f}^{(2)}$, we obtain
\begin{equation}
  P_2(\mathfrak{f})
  =
  \int (d\mathfrak{q})
  \bigl<
  \bigl(
    1 - i \mathfrak{q} \mathfrak{f}^{(2)}[\delta\alpha]
  \bigr)
  \exp
  i \mathfrak{q}
  \bigl(
    \mathfrak{f}
  - \mathfrak{f}^{(1)}[\delta\alpha]
  \bigr)
  \bigr> .
\end{equation}
The term with 1 gives the hard-core distribution function $P_0(\mathfrak{f})$, while the term with $\mathfrak{f}^{(2)}[\delta\alpha]$ provides corrections to it. According to Wick's theorem, there are two types of these corrections: one coming from $\mathfrak{q}_\mu \corr{\mathfrak{f}_\mu^{(2)}}$, and the other from the irreducible average $\mathfrak{q}_\mu\mathfrak{q}_\eta\mathfrak{q}_\nu \ccorr{\mathfrak{f}_\mu^{(1)}\mathfrak{f}_\eta^{(2)}\mathfrak{f}_\nu^{(1)}}$. The former vanishes as an integral of a total derivative [see Eq.\ \eqref{f-exp-2ord}], and the latter is calculated as
\be
  \ccorr{\mathfrak{f}_\mu^{(1)}\mathfrak{f}_\eta^{(2)}\mathfrak{f}_\nu^{(1)}}
  =
  %-
  2
  \zeta_\alpha^2
  I_{\mu\nu}^\eta ,
\ee
where
\be
  I_{\mu\nu}^\eta = 
  %-
  \int d\br \, d\br' \, 
  \mathfrak{b}_\mu(\br)
  \mathfrak{c}_\eta(\br,\br')
  \mathfrak{b}_\nu(\br'),
\label{I mu and I mu nu eta definition}
\ee
and the vector functions $\mathfrak{b}(\br)$ and $\mathfrak{c}(\br,\br')$ are defined in Eqs.\ \eqref{Def-b-c}.
Finally, transforming $\mathfrak{q}_\mu\mathfrak{q}_\eta\mathfrak{q}_\nu$ using integration by parts, we end up with \begin{equation}
\label{P-f-2ord}
    P_2(\mathfrak{f})
    =
   \left(1
   %+
   -
   \frac{\zeta_\alpha^2}{2}I_{\mu\nu}^\eta
\partial_{\mathfrak{f}_\mu}\partial_{\mathfrak{f}_\nu}
\partial_{\mathfrak{f}_\eta}
   \right)
    P_0(\mathfrak{f}) .
\end{equation}

Equation \eqref{P-f-2ord} is the main result of this Section, providing the leading correction to the hard-core distribution function. This correction arises from the vortex core's adjustment to the disorder potential. Since $P_0(\mathfrak{f})$ is simple Gaussian, the main challenge is to calculate $I_{\mu\nu}^\eta$. We also note 
that the form of $P_2(\mathfrak{f})$ automatically ensures that the distribution function remains normalized.

\subsection{Density of extrema}
\label{Sec-density-ext}

As discussed in Sec.\ \ref{SS:Disclaimer}, the kernel $\mathfrak{c}_\eta(\br,\br')$ can be correctly defined only once a procedure for selecting the vortex center is specified. This ambiguity is reflected in the dependence of $I_{\mu\nu}^\eta$ and, consequently, the distribution function $P_2(\mathfrak{f})$ on the particular choice 
of the regularized operator $\hat{L}_\text{reg}$.
The density of extrema is the only quantity that appears invariant under different vortex-centering prescriptions. 
Here we calculate it and demonstrate that the operator $\hat L^{-1}$ is indeed invertible on the resulting function class.

\subsubsection{Representation in terms of $\corr{M|L|N}$}

For clarity, we focus first on the density of minima. The corresponding results for maxima and saddle points are presented at the end of this section. Substituting the first correction to the distribution function \eqref{P-f-2ord} into Eq.\ \eqref{n-a-def} yields an expression for the change in the density of minima as a contraction of two tensors:
\begin{equation}
\label{def-delta-n-a}
    \delta n_\text{min}
    =
    %+
    -
    \frac{\zeta_\alpha^2}{2}I_{\mu\nu}^\eta S_{\mu\nu\eta},
\end{equation}
where
\be
\label{S-def}
     S_{\mu\nu\eta}
    =
    \int
    d^5\mathfrak{f} \,
    \delta(\bf)
    \left|\det H\right| \Theta_\text{min}\,
    \partial_{\mathfrak{f}_\mu}\partial_{\mathfrak{f}_\nu}
\partial_{\mathfrak{f}_\eta}
    P_0(\mathfrak{f}) .
\ee
The symmetric tensor $S_{\mu\nu\eta}$ is calculated in Appendix \ref{Appendix-S}.
Substituting Eqs.\ \eqref{S-s} and \eqref{s-res} into Eq.\ \eqref{def-delta-n-a}, we obtain $\delta n_\text{min}$ as a sum of two contributions, which differ by the number of indices in $I^\eta_{\mu\nu}$:
\be
\label{deltan1}
 \delta n_\text{min}
 =
 %+
 -
 \frac{\zeta_\alpha^{1/2}}{48\Delta_0^6}
 \frac{1}{\sqrt{\pi^3c_1^2c_2}}
 \left(
 \frac{8 c_2}{c_1\xi}\Sigma_4+\frac{\xi}{3}\Sigma_6
 \right) ,
\ee
where
\begin{multline}
\label{Sigma4-I}
  \Sigma_4 
  =
  (I_{y,xx}^y+I_{xx,y}^y+I_{y,y}^{xx})
  +
  (I_{x,yy}^x+I_{x,x}^{yy}+I_{yy,x}^x)
\\{}
  +
  (I_{x,x}^{xx}+I_{xx,x}^x+I_{x,xx}^x)
  +
  (I_{y,yy}^y+I_{y,y}^{yy}+I_{yy,y}^y)
\end{multline}
and
\begin{multline}
\label{Sigma6-I}
\Sigma_6
=
I_{xx,xx}^{xx}+I_{yy,yy}^{yy}
\\{}
+8(I_{xy,xy}^{xx}+I_{xx,xy}^{xy}+I_{xy,xx}^{xy}+I_{xy,xy}^{yy}+I_{yy,xy}^{xy}+I_{xy,yy}^{xy})
\hspace{10pt}
\\{}
-7(I_{xx,yy}^{xx}+I_{yy,xx}^{xx}+I_{xx,xx}^{yy}+I_{yy,xx}^{yy}+I_{xx,yy}^{yy}+I_{yy,yy}^{xx}) .
\end{multline}

The quantities $I^\eta_{\mu\nu}$, defined by the double integral in Eq.\ \eqref{I mu and I mu nu eta definition}, involve the factor $\mathfrak{c}_\eta(\br,\br')$, which contains either one ($\eta=x$ or $y$) or two ($\eta=xx$, $xy$, or $yy$) derivatives of $\hat{L}(\br,\br';\bR)$. 
Therefore, it is convenient to transfer the derivatives from $\mathfrak{c}_{\eta}$ to $\mathfrak{b}_{\mu}$ and $\mathfrak{b}_{\nu}$. To this end, we replace the derivatives $\hat\partial_\mu$ with respect to the vortex center $\bR$ in Eq.\ \eqref{I mu and I mu nu eta definition} by those with respect to $\br$ and $\br'$, using $(\partial_{R_i}+\partial_{r_i})f(\br-\bR)=0$, etc. Then we set $\bR=0$ and, integrating by parts, make $\hat{L}(\br,\br';\bR)$ free of derivatives.

The resulting expressions are conveniently presented in the bra-ket notations if we introduce a vector $|ij\dots l\rangle$ according to
\be
\label{ket-def}
  \langle\br|ij\dots l\rangle
  = 
  \hat{\Delta}_0(\br) \,
  \partial_i \partial_j \dots \partial_l|\Delta_0(\br)|^2 ,
\ee
and an associated matrix element of the fluctuation propagator:
\be
\label{MLN-def}
  \corr{M|L|N} 
  = 
%  -
  \int d\br \, d\br' \, 
  \corr{M|\br}
  \hat{L}(\br,\br')
  \corr{\br'|N} ,
\ee
where now $M$ and $N$ are arbitrary strings of $x$ and $y$.

Then we obtain for $I^\eta_{\mu\nu}$:
\begin{subequations}
\label{I-T}
\be
    I_{\mu\nu}^{i}=  (-1)^{|\mu|+|\nu|+1}
    %\sigma_{\mu\nu} 
    \left[ \corr{\mu i|L|\nu} + \corr{\mu|L|\nu i} \right] ,
    \label{J1-over-T}
\ee
\vspace{-10pt}
\begin{multline}
    I_{\mu\nu}^{ij} = (-1)^{|\mu|+|\nu|+1}
    %\sigma_{\mu\nu} 
    \left[ 
      \corr{\mu i j|L|\nu} + \corr{\mu i|L|\nu j}
    \right.
\\{}
    \left.  
       + \corr{\mu j|L|\nu i} + \corr{\mu|L|\nu i j} 
    \right],
    \label{J2-over-T}
\end{multline}
\end{subequations}
where $|M|$ is the length of the string $M$ (e.g., $|xy|=2$).

Using the symmetry of $\corr{M|L|N}$ with respect to arbitrary permutations of $x$ and $y$ in $M$ and $N$, we rewrite Eqs.\ \eqref{Sigma4-I} and \eqref{Sigma6-I} as
\begin{equation}
\label{Sigma4-T}
    \Sigma_4
    =
    %4\corr{xy|L|xy} - 4\corr{xx|L|yy}
    4\corr{x^2|L|y^2} - 4\corr{xy|L|xy}
\end{equation}
and
\begin{multline}
\label{Sigma6-T}
  \Sigma_6
  =
  4\langle 7 y^4  + 6 x^2y^2 - x^4 |L|x^2\rangle 
  - 64 \langle x^3y|L|xy\rangle
\\
  - 4\langle 3xy^2-x^3|L|3xy^2-x^3\rangle .
\end{multline}
Here we treat bras and kets as formal symbols for linear operations. Using bilinearity, any expression of this form first should be expanded as a linear combination of the basic matrix elements, and only then their definition \eqref{MLN-def} should be applied.

\subsubsection{Zero overlap with the zero mode}
\label{independence part}

The two components of $|N\rangle$ defined in Eq.\ \eqref{ket-def} are complex conjugates:
\be
\label{hat-Phi}
  \hat\Phi_N(\br)
  =
  \corr{\br|N}
  = 
  \begin{pmatrix}
    \Phi_N(\br) \\ \Phi_N^*(\br)
  \end{pmatrix} .
\ee
A convenient approach for both analysis and computation is to expand its first component in angular momentum eigenstates:
\be
\label{J-momenta}
  \Phi_N(\br)
  = 
  \sum_m \Phi_N^m(r) e^{i(m+1)\vp}
  .
\ee
For a given $N$, the sum contains momenta from $-|N|$ to $|N|$ in steps of 2 ($|N|+1$ terms in total). 
The operator $\hat L^{-1}$ in this basis couples only momenta $m$ and $-m$, see Appendix \ref{Appendix-J-I}.

The zero modes $|0_{x,y}\rangle$, with the wave functions \eqref{def-zero-mode}, have angular momenta $\pm1$. To prove they do not appear in calculating $\delta n_\text{min}$, 
it is sufficient to verify that their overlap with the bras and kets appearing in $\Sigma_4$ and $\Sigma_6$ is zero.

Equation \eqref{Sigma4-T} for $\Sigma_4$ contains only $\corr{M|L|N}$ with even $|M|$ and $|N|$. The corresponding momenta are even and therefore have zero overlap with $|0_{x,y}\rangle$. The same also holds for the first line in Eq.\ \eqref{Sigma6-T} for $\Sigma_6$. 
Its second line involves an odd-momentum expression, but the identity $3\partial_x\partial_y^2-\partial_x^3=-4(\partial_z^3+\partial_{\overline{z}}^3)$ with $z=x+iy$ and $\overline{z}=x-iy$ restricts $|3xy^2-x^3\rangle$ to momenta $\pm3$, which likewise yields zero overlap with the zero modes.

The vanishing overlaps $\corr{0_{x,y}|N}=0$ for all states $|N\rangle$ in $\Sigma_4$ and $\Sigma_6$ provide an independent check of consistency. They demonstrate that the areal density $n_\text{min}$ is an invariant of the pinning potential, unaffected by the choice of vortex-centering prescription. Practically, this justifies the use of the original operator $\hat{L}$ instead of its regularized counterpart $\hat{L}_\text{reg}$ introduced in Sec.\ \ref{SS:Disclaimer}.

\subsubsection{Final result}

The lack of an analytic expression for $\hat{L}(\br,\br';\bR)$ means $\corr{M|L|N}$ must be evaluated numerically. 
Incorporating the results from Appendix \ref{Appendix-T-calc} into Eqs.\ \eqref{Sigma4-T} and \eqref{Sigma6-T}, and then into Eq.\ \eqref{deltan1}, we obtain the correction to the areal density of minima due to core deformation:
\begin{equation}
  \delta n_\text{min}
  =
  - \frac{0.0099}{\xi^2} \frac{\sqrt{\A}}{\alpha \xi} 
  =
  - \frac{0.0351}{\xi^2} \sqrt{\frac{\tau_*}{\tau}} ,
\end{equation}
where $\tau_*$ is the dimensionless inhomogeneity strength introduced in Eq.\ \eqref{tau*}. 

Repeating the same steps for maxima and saddles, and adding the hard-core contributions (see Sec.\ \ref{Sec:Density of extrema}), we arrive at the main result of our research:
\begin{subequations}
\label{n-final}
\begin{align}
\label{n-min-final}
  n_\text{min} & = 
  \frac{0.0276}{\xi^2}
  \left[ 
    1 
    - 1.27 \sqrt{\frac{\tau_*}{\tau}} 
    + O\Big(\frac{\tau_*}{\tau}\Bigr)
  \right] ,
\\
  n_\text{max} & = 
  \frac{0.0276}{\xi^2}
  \left[ 
    1 
    + 1.27 \sqrt{\frac{\tau_*}{\tau}} 
    + O\Big(\frac{\tau_*}{\tau}\Bigr)
  \right] ,
\\
  n_\text{sad} & = 
  \frac{0.0552}{\xi^2}
  \left[ 
    1 
    + O\Big(\frac{\tau_*}{\tau}\Bigr)
  \right] .
\end{align}
\end{subequations}

Equations \eqref{n-final} give the mean areal densities of the different extremum types of the vortex pinning potential. The first term in the brackets is the hard-core result, while the second term provides the correction due to vortex core distortion by impurities. 
We see that allowing the vortex core to adapt to the quenched disorder breaks the symmetry between the number of minima and maxima, reducing the former and enhancing the latter. Meanwhile, the number of saddle points is not modified. The equality $n_\text{min}+n_\text{max}=n_\text{sad}$, which we have verified perturbatively, is a direct manifestation of Poincar\'e-Hopf index theorem, a cornerstone of differential topology.

Finally, we emphasize that our results \eqref{n-final} are nonperturbative in inhomogeneity strength $\A$.

\section{Summary and conclusion}
\label{S:Conclusion}

This work re-examines vortex pinning in superconducting films with short-scale inhomogeneities. Employing the random-temperature Ginzburg-Landau framework, which provides a universal model of frozen-in disorder close to $T_c$ \cite{Larkin-2,SF2013,Zuev}, we relate the vortex pinning landscape $E(\bR)$ to the spatially varying coefficient $\delta\alpha(\br)$.

For sufficiently weak disorder, when the vortex core can be considered as rigid, the modulation of $E(\bR)$ is a linear functional of $\delta\alpha(\br)$, thereby inheriting its Gaussian statistics. In this limit, we calculate the joint distribution function of the local force on the vortex and its Hessian matrix. Our results are in agreement with Refs.\ \cite{Larkin-1,Geshkenbein2022} and extend them, providing the mean density of extrema (minima, maxima, and saddle points) and the associated local curvature distribution $P_a(\kappa_1,\kappa_2)$.

We also consider vortex pinning beyond the hard-core approximation by allowing the order parameter to relax in response to the random potential. In this case, the ambiguity in defining the vortex center makes the pinning landscape $E(\bR)$ prescription-dependent. However, the \emph{number} of its \emph{extrema} remains well-defined. Equations~\eqref{n-final} give the leading correction to the areal density of extrema due to the core softening. The adaptation of the vortex core to a disordered environment reduces the number of available stable configurations. The relative magnitude of this reduction scales as $\sqrt{\tau_*/\tau}$, where $\tau$ is the reduced temperature and $\tau_*$ is the ``dirty Ginzburg'' parameter (at $\tau\lesssim\tau_*$, the superconducting state becomes strongly inhomogeneous) \cite{Zuev,SF2005}.

We find that near $T_c$, the characteristic distance between adjacent pinning minima is $\Delta r\sim 6\xi$. 
A large numerical factor in $\Delta r/\xi$ is relevant for understanding individual vortex creep and its effect on the vortex lattice.
An important open question is to extend our analysis and trace the evolution of $\Delta r/\xi$ to lower temperatures.

The developed theory finds a natural application in homogeneously disordered thin films of the NbN family, key materials for modern superconducting devices such as single-photon detectors \cite{Goltsman,Hadfield} and quantum phase-slip circuits \cite{Mooij,Astafiev}. For these technologies, understanding vortex pinning and motion \cite{Pratap2019} is vital for device optimization.

\acknowledgments

The authors are grateful to V. B. Geshkenbein for useful discussions. 
This research was supported by the Russian Science Foundation under Grant No. 23-12-00297.

\appendix

\section{GL functional disturbed by $\delta\alpha(\br)$}
\label{Appendix-A}

In this Appendix, we derive the first- and second-order corrections to the energy of a saddle-point solution $\Delta_0(\br)$ disturbed by the field $\delta\alpha(\br)$, which enters linearly in a generic GL-type functional $F[\Delta,\Delta^*;\alpha]$. The obtained expressions should be applied to the vortex solution \eqref{vortex-g} for the free energy \eqref{def-GL-Functional}.

Writing the order parameter as $\Delta(\br) = \Delta_0(\br) + \Delta_1(\br) + \Delta_2(\br) + \dots$, with $\Delta_n$ scaling as $(\delta\alpha)^n$, we solve the saddle-point equation
$\delta F/\delta\Delta^*(\br)=0$ and get an integro-differential equation for the first-order correction:
\be
\label{Def-Delta-1-eq}
  \int \! d\br'
  \big[
    F_{\Delta^*_{\br}\Delta_{\br'}}\Delta_1(\br')
    + F_{\Delta^*_{\br}\Delta^*_{\br'}}\Delta_1^*(\br')
    + F_{\Delta^*_{\br}\alpha_{\br'}}\delta\alpha(\br')
  \big]
  =
  0 ,
\ee
where $F$ with superscripts denotes a functional derivative of $F[\Delta,\Delta^*;\alpha]$ evaluated at the bare saddle-point solution $\Delta_0$ and at $\delta\alpha=0$.

Then we perform a routine perturbative expansion and utilize the fact that $\Delta_0(\br)$ is the saddle-point of the unperturbed functional, ensuring $F_{\Delta}=F_{\Delta^*}=0$. We also use the linearity of the functional in $\alpha$. As a result, the first- and second-order corrections to the saddle-point energy simplify to
\begin{gather}
\label{F-1-expanded}
    F_1
    =
    \int d\br \,
    F_{\alpha_{\br}} \delta\alpha(\br) ,
\\
    \label{F-2-expanded}
    F_2
    =
    \frac{1}{2}\int d\br\, d\br' \,
    \delta\alpha(\br)
    \big[
    F_{\alpha_{\br}\Delta_{\br'}}\Delta_1(\br')
    + F_{\alpha_{\br}\Delta_{\br'}^*}\Delta_1^*(\br')
    \big] ,
\end{gather}
where $\Delta_1$ is determined by Eq.\ \eqref{Def-Delta-1-eq}.

For the GL functional \eqref{def-GL-Functional},
$F_{\alpha_{\br}} = - |\Delta_0(\br)|^2$, hence
$F_{\alpha_{\br}\Delta_{\br'}^*} = -\delta(\br-\br')\,\Delta(\br')$ and
$F_{\alpha_{\br}\Delta_{\br'}} = -\delta(\br-\br')\,\Delta^*(\br')$.
So we arrive at Eq.\ \eqref{E12}.

If a constraint $W[\Delta,\Delta^*]$ is added to the functional $F[\Delta,\Delta^*]$, it must be incorporated via a Lagrange multiplier. One can verify that that Eqs.\ \eqref{F-1-expanded} and \eqref{F-2-expanded} remain unchanged, and only the first-order correction $\Delta_1$ should modified according to Eq. \eqref{Delta-1-eq-with-projections}.

\section{Handling zero modes}
\label{Appendix-zero-mode}

As discussed in Sec.\ \ref{SS:Disclaimer}, the divergent normalization of the zero mode prevents a straightforward projection of the right-hand side in Eq.\ \eqref{Delta-1-eq} onto the space orthogonal to the zero modes. 
To resolve this, we impose an auxiliary condition that fixes the vortex center through the minimization of a chosen functional [Eq. \eqref{W-exp}]. Incorporating this condition via Lagrange multipliers $\Lambda_x$ and $\Lambda_y$ modifies the original functional \eqref{def-GL-Functional} to
\begin{equation}
    \label{def-F-eff}
    \Feff [\Delta(\br),\Lambda; \bR] 
    = 
    \mathcal{F} [\Delta(\br)] + \Lambda_i W_i[\Delta(\br);\bR],
\end{equation}
where
\begin{equation}
    W_i[\Delta(\br);\bR]
  =
  \int d\br\, \Upsilon(|\br-\bR|) \frac{\partial}{\partial R_i} |\Delta(\br)-\Delta_{0,\bR}(\br)|^2 .
\end{equation}

Minimizing the functional \eqref{def-F-eff} and expanding to the first order in $\delta\alpha$, we obtain the linear system for $\Delta_1(\br)$ and $\Lambda_i$:
\begin{subequations}
\label{Delta-1-eq-with-projections}
\begin{gather}
    \label{Delta1-diff-eq-with-projections}
    \hspace{6mm}
    \hat{L}^{-1} \hat{\Delta}_1 = \delta\alpha(\br) \hat{\Delta}_0 - \Upsilon(\br) \lambda_i \partial_i \hat{\Delta}_0 ,
\\
    \int d\br \, \Upsilon(\br) \hat{\Delta}_1^\dagger \partial_i \hat{\Delta}_0 = 0 .
\end{gather}
\end{subequations}
Due to the presence of the last term in Eq.\ \eqref{Delta1-diff-eq-with-projections}, the Lagrange multipliers $\lambda_i$ can be chosen such that the right-hand side is orthogonal to zero modes. This renders the system \eqref{Delta-1-eq-with-projections} well-posed: its solution exists and is unique.

In addition to translational invariance, the unperturbed system has a continuous U(1) symmetry corresponding to global phase rotations, along with the corresponding zero mode $|0_\vp\rangle$. However, the orthogonality condition for this mode is satisfied automatically, and we need not make any modifications to account for it.

\section{Computation of $\corr{M|L|N}$}
\label{Appendix-J-I}

For computational purposes, we write the objects defined in Eq.\ \eqref{MLN-def} as
\begin{equation}
    \corr{M|L|N} = \int d\br \, \hat\Phi_M^\dagger(\br)  \hat{J}_N(\br) ,
\end{equation}
where
$\hat\Phi_{N}(\br) =\langle\br|N\rangle$ [see Eq.\ \eqref{ket-def}] and
\begin{equation}
\label{J-def}
    \hat{J}_N(\br)
    = \int d\br^\prime\,
  \hat{L}(\br,\br')
  \hat\Phi_N(\br').
\end{equation}

The main difficulty in calculating $\corr{M|L|N}$ comes from finding $\hat{J}_N(\br)$. Since the kernel $\hat{L}(\br,\br')$ is unknown, it is better to write Eq.\ \eqref{J-def} as a differential equation
\begin{equation}
\label{J-eq-def}
    \hat{L}^{-1}\hat{J}_N(\br)
    =
    \hat\Phi_N(\br).
\end{equation}
The proposed scheme of the zero mode elimination modifies this equation similar to the way \eqref{Delta-1-eq} becomes \eqref{Delta-1-eq-with-projections}:%
\begin{subequations}
\begin{gather}
\label{eq-J-Phi}
    \hspace{8mm}
    \hat{L}^{-1} \hat{J}_N 
    = 
    \hat\Phi_N(\br)  
    - \Upsilon(\br) \lambda_i \hat\psi_i ,
\\
    \int d\br \, \Upsilon(\br) \hat{J}_N^\dagger \hat\psi_i = 0 .
\label{eq-J-lambda}
\end{gather}
\end{subequations}

We solve this system numerically as follows. First, we determine $\lambda_i$ by requiring the right-hand side of Eq.~\eqref{eq-J-Phi} to be orthogonal to the zero modes, ensuring a solution exists. Once $\lambda_i$ are known, we compute a particular numerical solution $\hat{J}_N^{(p)}$. The general solution is then $\hat{J}_N = \hat{J}_N^{(p)} + a_i \hat{\psi}_i$. Finally, the coefficients $a_i$ are obtained by solving Eq.\ \eqref{eq-J-lambda}, which is linear in $a_i$.

The numerical solution of a 2D differential equation \eqref{eq-J-Phi} is significantly simplified in the angular momentum representation \eqref{J-momenta}, since $\hat L^{-1}$ couples only $J^m_N$ and $(J_N^{-m})^*$:
\be
\label{J-eq-1d}
  \hat L_m^{-1}
  \begin{pmatrix} J^m_N \\ J_N^{-m,*} \end{pmatrix} 
  = 
  \begin{pmatrix} 
    \Phi_N^m \\  \Phi_N^{-m,*}
  \end{pmatrix} 
  - \lambda_i \Upsilon(r) \begin{pmatrix} 
    \psi_i^{m} \\ 
     \psi_i^{-m,*} 
  \end{pmatrix}
  ,
\ee
where we follow the notation of Eq.\ \eqref{hat-Phi}, with an object without a hat being the first element of the corresponding vector.
In the momentum basis,
\be
  \hat L_m^{-1}
  =
  \big[ - \gamma \Delta_r^{m+\sigma_3} - \alpha + 2 \beta|\Delta_0|^2 \big] \sigma_0
  + \beta|\Delta_0|^2 \sigma_1 ,
\ee
where $\sigma_i$ are the Pauli matrices,
and $\Delta_r^m$ is the radial part of the 2D Laplace operator corresponding to the angular momentum $m$:
\begin{equation}
    \Delta_r^m = \partial_r^2 + \frac{1}{r} \partial_r - \frac{m^2}{r^2}.
\end{equation}

For any given string $N$, $\hat\Phi_N$ contains a finite number of non-zero angular harmonics. Therefore, computing $\hat J_N(\br)$ (and thus $\corr{M|L|N}$) reduces to solving a set of few coupled one-dimensional equations \eqref{J-eq-1d}.

\section{Calculation of $S_{\mu\nu\eta}$}
\label{Appendix-S}

Here we calculate nonzero elements of $S_{\mu\nu\eta}$, which is a symmetric rank-3 tensor defined in Eq.\ \eqref{S-def} via an integral over 5-vector $\mathfrak{f}$. Symmetry relations dictate that there are four nontrivial elements: $S_{x,x,xx}$, $S_{xx,xx,xx}$, $S_{xx,xx,yy}$, and $S_{xy,xy,xx}$. All other nonzero elements can be obtained from those by applying the transformations (i) $x\leftrightarrow y$ and (ii) $xx\leftrightarrow yy$.

Switching to dimensionless variables $t_i=f_i/\sqrt{c_1\A\Delta_0^4}$ and  $t_a=\sqrt{\xi^2/2c_2\A\Delta_0^4}\, h_a$, we obtain
\begin{equation}
\label{S-s}
  S_{\mu\nu\eta}
  = 
  \frac{1}{(2\pi)^{5/2}}\frac{1}{\sqrt{2c_1^2c_2}}
  \left(\frac{2c_2}{c_1}\right)^{\omega/2}
  \frac{\xi^{1-\omega}}{\zeta_\alpha^{3/2}\Delta_0^6}
  s_{\mu\nu\eta} ,
\end{equation}
where $\omega=6-|\mu|-|\nu|-|\eta|$ and
\begin{multline}
     s_{\mu\nu\eta}
    =
    \int
    d^5t \,
    \delta(t_i)
    |t_{xx}t_{yy}-t_{xy}^2| \, \Theta_\text{min}
\\{}
\times
  \partial_{t_\mu}\partial_{t_\nu}\partial_{t_\eta}
  \exp\bigl(-t_x^2/2-t_y^2/2-t_a C^{-1}_{ab}t_b\bigr) ,
\end{multline}
with $\Theta_\text{min}=\theta(t_{xx})\theta(t_{xx}t_{yy}-t^2_{xy})$.
This integral is evaluated following the procedure described in Sec.\ \ref{Sec:Density of extrema}: using the Fourier transform of the absolute value [Eq.\ \eqref{mod-FT}] and calculating the remaining Gaussian integral over $t$. As a result, we arrive at
\begin{equation}
  s_{\mu\nu\eta}
  =
  -
  \sqrt{\frac23} \, \pi
  \int\frac{d\lambda}{2\pi}
  \frac{K_{\mu\nu\eta}}{(\lambda-i\epsilon)^{2}} ,
\end{equation}
where nonzero terms are given by
\begin{subequations}
\begin{align}
  K_{1,1,11}
  & =
 \frac{1}{\sqrt{1+i \lambda } (1-2i \lambda)},
\\
  K_{11,11,11}
  & =
  \frac{2-7i\lambda}{3 \sqrt{1+i \lambda } (1-2i\lambda)^2},
\\
  K_{11,11,22}
  & =
  -
  \frac{i\lambda }{\sqrt{1+i \lambda } (1-2i\lambda)^2} ,
\\
  K_{12,12,11}
  & =
  \frac{2i\lambda}{(1+i\lambda)^{3/2} (1-2i \lambda)} .
\end{align}
\end{subequations}
Closing the contour in the lower half-plane and taking the residue at $\lambda=-i/2$, we immediately obtain
\be
\label{s-res}
{
  s_{1,1,11} = 4 \pi/3 ,
\quad 
  s_{11,11,11}=\pi/9 ,
} 
\atop
{
  s_{11,11,22}=-7\pi/9 ,
\quad
  s_{12,12,11}=8\pi/9 .
}
\ee

\section{numerical expressions for $\corr{M|L|N}$}
\label{Appendix-T-calc}

In this Appendix, we present the numerical results for $\corr{M|L|N}$ that enter $\Sigma_4$ [Eq.\ \eqref{Sigma4-T}] and $\Sigma_6$ [Eq.\ \eqref{Sigma6-T}]. First we separate dimensional factors by writing
\begin{equation}
\label{L-l}
  \corr{M|L|N}
  =
  \Delta_0^6\xi^{2-|M|-|N|} \,
  \corr{M|l|N} .
\end{equation}
The matrix elements $\corr{M|l|N}$ are calculated numerically by solving a finite number of one-dimensional $2\times2$ matrix differential equations \eqref{J-eq-1d}. Thereby we obtain
\begin{subequations}
\label{overlap-numetic}
\begin{align}
    &\langle xy|l|xy\rangle  =0.227, \\
    &\langle y^2|l|x^2\rangle  =-0.161, \\
    &\langle x^4 |l|x^2\rangle  =-0.391, \\
    &\langle x^2y^2|l|x^2\rangle  =-0.0086, \\
    &\langle y^4|l|x^2\rangle  =0.340, \\
    &\langle x^3y|l|xy\rangle  =-0.183, \\
    &\langle 3xy^2-x^3|l|3xy^2-x^3\rangle  =1.352.
\end{align}
\end{subequations}


\begin{thebibliography}{99}

\bibitem{Blatter1994}
G. Blatter, M. V. Feigel'man, V. B. Geshkenbein, A. I. Larkin, and V. M. Vinokur,
Vortices in high-temperature superconductors,
Rev. Mod. Phys. \textbf{66}, 1125 (1994).
%DOI: https://doi.org/10.1103/RevModPhys.66.1125 

\bibitem{BS}
J. Bardeen and M. Steven,
Theory of the motion of vortices in superconductors, 
Phys. Rev. \textbf{140}, A1197 (1965).


\bibitem{Si}
W. Si, S. J. Han, X. Shi, S. N. Ehrlich, J. Jaroszynski, A. Goyal, and Q. Li, High current superconductivity in FeSe$_{0.5}$Te$_{0.5}$-coated conductors at
30 Tesla, Nat. Commun. \textbf{4}, 1347 (2013).

\bibitem{Devoret}
M. H. Devoret and R. J. Schoelkopf,  Superconducting circuits for quantum
information: an outlook, Science \textbf{339}, 1169 (2013).

\bibitem{Welp}
U. Welp, K. Kadowaki, and R. Kleiner, Superconducting emitters of THz
radiation, Nat. Photon. \textbf{7}, 702 (2013).

\bibitem{Gurevich}
A. Gurevich and G. Ciovati, Dynamics of vortex penetration, jumpwise
instabilities, and nonlinear surface resistance of type-II superconductors in strong rf fields, Phys. Rev. B \textbf{77}, 104501 (2008).


\bibitem{Larkin-0}
A. I. Larkin, Effect of inhomogeneities on the structure of the mixed state of superconductors, 
Zh. Eksp. Teor. Fiz. \textbf{58}, 1466 (1970) 
[Sov. Phys. JETP \textbf{31}, 784 (1970)].

\bibitem{CollectivePinning}
A. I. Larkin and Yu. Ovchinnikov, 
Pinning in type II superconductors,
J. Low Temp. Phys. \textbf{34}, 409 (1979).

%\bibitem{CollectivePinning2D}
%P. H. Kes and C. C. Tsuei,
%Two-dimensional collective flux pinning, defects, and structural %relaxation in amorphous superconducting films,
%Phys. Rev. B \textbf{28}, 5126 (1983).
%DOI: https://doi.org/10.1103/PhysRevB.28.5126 

\bibitem{FS97}
M. V. Feigel'man and M. A. Skvortsov,
Anomalous flux-flow dynamics in layered type-II superconductors at low temperatures,
Phys. Rev. Lett. \textbf{78}, 2640 (1997).

\bibitem{LO98}
A. I. Larkin and Yu. N. Ovchinnikov,
Resistance of layered superclean superconductors at low temperatures, Phys. Rev. B \textbf{57}, 5457 (1998).

\bibitem{Koulakov}
A. A. Koulakov and A. I. Larkin, Vortex density of states and absorption in clean layered superconductors, Phys. Rev. B \textbf{60}, 14597 (1999).

\bibitem{SKF98}
M. A. Skvortsov, V. E. Kravtsov, and M. V. Feigel'man,
Level statistics inside the core of a superconductive vortex, 
Pis'ma v ZhETF \textbf{68}, 78 (1998) [JETP Lett. \textbf{68}, 84 (1998)].

\bibitem{BHL99}
E. Br\'ezin, S. Hikami, and A. I. Larkin, Level statistics inside the vortex of a superconductor and symplectic random-matrix theory in an external source, Phys. Rev. B \textbf{60}, 3589 (1999).

\bibitem{Fujita}
A. Fujita, Level statistics for the quasiparticle spectra inside a two-dimensional vortex core with impurities, Phys. Rev. B \textbf{62}, 15190 (2000).

\bibitem{Melnikov09}
A. S. Mel’nikov, A. V. Samokhvalov, and M. N. Zubarev,
Electronic structure of vortices pinned by columnar defects,
Phys. Rev. B \textbf{79}, 134529 (2009).

\bibitem{Melnikov2020}
A. V. Samokhvalov, V. D. Plastovets, and A. S. Mel'nikov, Topological transitions in electronic spectra: Crossover between Abrikosov and Josephson vortices,
Phys. Rev. B \textbf{102}, 174501 (2020).

\bibitem{Khodaeva2022}
U. E. Khodaeva and M. A. Skvortsov,
Vortex core near planar defects in a clean layered superconductor,
Phys. Rev. B \textbf{105}, 134504 (2022).


\bibitem{Suderow}
H. Suderow, I. Guillam\'on, J. G. Rodrigo, and S. Vieira,
Imaging superconducting vortex cores and lattices with a scanning tunneling microscope, 
Supercond. Sci. Technol. \textbf{27}, 063001 (2014).

\bibitem{Pratap2023}
R. Duhan, S. Sengupta, R. Tomar, S. Basistha, V. Bagwe, C. Dasgupta, and P. Raychaudhuri,
Structure and dynamics of a pinned vortex liquid in a superconducting $\alpha$-Re$_6$Zr thin film,
Phys. Rev. B \textbf{108}, L180503 (2023).
%DOI: https://doi.org/10.1103/PhysRevB.108.L180503 

\bibitem{Zeldov-1}
L. Embon, Y. Anahory, A. Suhov, D. Halbertal, J. Cuppens, A. Yakovenko, A. Uri, Y. Myasoedov, M. L. Rappaport, M. E. Huber, A. Gurevich, and E. Zeldov, Probing dynamics and pinning of single vortices in superconductors at nanometer scales, Sci. Rep. \textbf{5}, 7598 (2015).

\bibitem{Zeldov-2}
L. Embon, Y. Anahory, \v{Z}. L. Jeli\'c, E. O. Lachman,
Y. Myasoedov, M. E. Huber, G. P. Mikitik, A. V. Silhanek, M. V. Milo\v{s}evi\'c, A. Gurevich, and E. Zeldov, Imaging of super-fast dynamics and flow instabilities of superconducting vortices, 
Nat. Commun. \textbf{8}, 85 (2017).

\bibitem{Zeldov2019}
K. Bagani, J. Sarkar, A. Uri, M. L. Rappaport,
M. E. Huber, E. Zeldov, and Y. Myasoedov, Sputtered Mo$_{66}$Re$_{34}$ SQUID-on-Tip for high-field magnetic and thermal nanoimaging, Phys. Rev. Appl. \textbf{12}, 044062 (2019).

\bibitem{Poggio}
L. Ceccarelli, D. Vasyukov, M. Wyss, G. Romagnoli,
N. Rossi, L. Moser, and M. Poggio, Imaging pinning
and expulsion of individual superconducting vortices in
amorphous MoSi thin films, Phys. Rev. B \textbf{100}, 104504
(2019).

\bibitem{Weber2025}
T. Weber \emph{et al.},
Advanced SQUID-on-lever scanning probe for high-sensitivity magnetic microscopy with sub-100-nm spatial resolution, Phys. Rev. Appl. \textbf{24}, 054041 (2025).
%more than 10, less than 20 autors

\bibitem{Thiel2016}
L. Thiel, D. Rohner, M. Ganzhorn, P. Appel, E. Neu,
B. M\"uller, R. Kleiner, D. Koelle, and P. Maletinsky, Quantitative nanoscale vortex imaging using a cryogenic quantum magnetometer, 
Nat. Nanotechnol. \textbf{11}, 677 (2016).

\bibitem{Foltyn2024}
M. Foltyn, K. Norowski, A. Savin, and M. Zgirski, Quantum thermodynamics with a single superconducting vortex, Sci. Adv. \textbf{10}, eado4032 (2024).

\bibitem{Bray2007}
A. J. Bray and D. S. Dean,
Statistics of critical points of Gaussian fields on large-Dimensional spaces,
Phys. Rev. Lett. \textbf{98}, 150201 (2007).

\bibitem{Yan2018}
Y. V. Fyodorov and P. L. Doussal,
Hessian spectrum at the global minimum of high-dimensional random landscapes,
J. Phys. A: Math. Theor. \textbf{51}, 474002 (2018).

\bibitem{Geshkenbein2022}
R. Willa, V. B. Geshkenbein, and G. Blatter,
Hessian characterization of the pinning landscape in a type-II superconductor,
Phys. Rev. B \textbf{105}, 144504 (2022).

\bibitem{Geshkenbein2023}
F. Gaggioli, G. Blatter, M. Buchacek, and V. B. Geshkenbein,
Strong pinning transition with arbitrary defect potentials,
Phys. Rev. Res. \textbf{5}, 033098 (2023).

\bibitem{Larkin-1}
A. I. Larkin and Yu. N. Ovchinnikov,
Influence of inho\-mo\-geneities on superconductor properties,
Zh. Eksp. Teor. Fiz. \textbf{61}, 1221 (1971)
[Sov. Phys. JETP \textbf{34}, 651 (1972)].

\bibitem{Zuev}
M. A. Skvortsov, O. B. Zuev, and D. I. Fazlizhanova, 
Supercurrent flow in inhomogeneous superconductors,
Phys. Rev. B \textbf{111}, 144510 (2025).
%arXiv:2412.00203.

\bibitem{Gorkov}
L. P. Gor’kov, Microscopic derivation of the Ginzburg-Landau equations in the theory of superconductivity,
Zh. Eksp. Teor. Fiz.  \textbf{36}, 1918 (1959) [Sov. Phys. JETP \textbf{9}, 1364
(1959)].

\bibitem{SF2005}
M. A. Skvortsov and M. V. Feigel’man,
Superconductivity in disordered thin films: Giant mesoscopic fluctuations,
Phys. Rev. Lett. \textbf{95}, 057002 (2005).
%DOI: https://doi.org/10.1103/PhysRevLett.95.057002 

\bibitem{Pearl}
J. Pearl, Current distribution in superconducting films carrying quantized fluxoids, Appl. Phys. Lett. \textbf{5}, 65 (1964).

\bibitem{Abrikosov}
A. A. Abrikosov, On the magnetic properties of superconductors of the second group, Zh. Eksp. Teor. Fiz.  \textbf{32}, 1442 (1957) [Sov. Phys. JETP \textbf{5}, 1174
(1957)].

\bibitem{Rice1}
S. O. Rice, Mathematical analysis of random noise, The Bell Sys. Tech. J. 
%The Bell System Technical Journal 
\textbf{23}, 282 (1944).

\bibitem{Rice2}
S. O. Rice, Mathematical analysis of random noise, The Bell Sys. Tech. J. 
%The Bell System Technical Journal 
\textbf{24}, 46 (1945).

\bibitem{Zittartz}
J. Zittartz and J. S. Langer,
Theory of bound states in a random potential,
Phys. Rev. \textbf{148}, 741 (1966).

\bibitem{Larkin-2}
A. I. Larkin and Yu. N. Ovchinnikov, 
Density of states in inhomogeneous superconductors, 
Zh. Eksp. Teor. Fiz. \textbf{61}, 2147 (1971)
[Sov. Phys. JETP \textbf{34}, 1144 (1972)].

\bibitem{SF2013}
M. A. Skvortsov and M. V. Feigel'man,
Subgap states in disordered superconductors,
Zh. Eksp. Teor. Fiz. \textbf{144}, 560 (2013)
[JETP 117, \textbf{487} (2013)].



\bibitem{Goltsman}
G. N. Gol'tsman, O. Okunev, G. Chulkova, A. Lipatov,
A. Semenov, K. Smirnov, B. Voronov, A. Dzardanov, C.
Williams, and R. Sobolewski, Picosecond superconducting single-photon optical detector, Appl. Phys. Lett. \textbf{79}, 705 (2001).

\bibitem{Hadfield}
C. M. Natarajan, M. G. Tanner, and R. H. Hadfield, Superconducting nanowire single-photon detectors: physics and applications, 
Supercond. Sci. Technol. \textbf{25}, 063001 (2012).

\bibitem{Mooij}
J. E. Mooij and Y. V. Nazarov, 
Superconducting na\-no\-wi\-res as quantum phase-slip junctions,
Nat. Phys. \textbf{2}, 169 (2006).

\bibitem{Astafiev}
R. S. Shaikhaidarov, K. H. Kim, J. W. Dunstan, I. V. Antonov, S. Linzen, M. Ziegler, D. S. Golubev, V. N. Antonov, E. V. Il’ichev, and O. V. Astafiev,
Quantized current steps due to the a.c.\ coherent quantum phase-slip effect,
Nature \textbf{608}, 45 (2022).

\bibitem{Pratap2019}
H. K. Kundu, K. R. Amin, J. Jesudasan, P. Raychaudhuri, S. Mukerjee, and A. Bid,
Effect of dimensionality on the vortex dynamics in a type-II superconductor,
Phys. Rev. B \textbf{100}, 174501 (2019).
%DOI: https://doi.org/10.1103/PhysRevB.100.174501 







\end{thebibliography}
\end{document}